\documentstyle[preprint,aps]{revtex}

\begin{document}
\draft
\title{Quantum teleportation with squeezed vacuum states}
\author{G.J. Milburn~$^*$ and S. L. Braunstein~$^\dagger$}
\address{$^*$~Department of Physics,
The University of Queensland,
QLD 4072 Australia.\\
$^\dagger$~SEECS, University of Wales, Bangor LL57 1UT, UK}
\date{\today}
\maketitle

\begin{abstract}
We show how the partial entanglement inherent in a two mode squeezed vacuum state admits two different teleportation
protocols. These two protocols refer to the different kinds of joint measurements that may be made by the sender. One protocol
is the recently implemented quadrature phase approach of Braunstein and Kimble[Phys. Rev. Lett.{\bf 80}, 869 (1998)]. The other
is based on recognising that a two mode squeezed vacuum state is also entangled with respect to photon number difference and
phase sum. We show that this protocol can also realise teleportation, however limitations can arise due to the fact that the
photon number spectrum is bounded from below by zero. Our examples show that {\em a given entanglement resource may admit more
than a single teleportation protocol and the question then arises as to what is the optimum protocol in the general case.}  
\end{abstract}
\newpage

\section{Introduction}

One of the central results in the rapidly developing field of quantum information theory is the possibility of perfectly
transferring an unknown quantum state from a target system at the sender's location, A,  to another identical system at the
receiver's location, B. This is called teleportation and requires that the sender and receiver share a maximally entangled
state, and further, that they can communicate via a classical channel.   The
original proposal of Bennett et al\cite{Bennett93} was posed in terms of systems with a two dimensional Hilbert space
(qubits\cite{Schumaker}). However recently Furasawa et al.\cite{Furasawa98}, using a proposal of Braunstein and
Kimble\cite{Braunstein98}, have demonstrated that the method can also be applied to entangled systems with an infinite
dimensional Hilbert space, specifically for harmonic oscillator states.  In that work, a coherent state was teleported using an
entanglement resource that consisted of a two mode squeezed vacuum state. The joint measurements required for teleportation
are joint quadrature phase on the target system and that part of the entangled resource shared by the receiver. The essential
feature exploited in the scheme of Furasawa et al is the well know fact that a two mode   squeezed vacuum state is an
approximation to an EPR (Einstein-Podolsky-Rosen) state, which had previously been shown by Vaidman\cite{vaidman94} to enable
teleportation of continuous observables.  However a squeezed vacuum state is also (imperfectly) entangled in number and phase.
Can this entanglement be used as a teleportation resource as well ?

In this paper we
show that by making joint number and phase measurements this entanglement can also be used for
teleportation. However because the number operator is bounded from below, there are limits on the ability to
teleport a quantum state by this protocol. 

Suppose that at some prior time a two mode squeezed vacuum state is generated and that one mode is open to local operations
and measurements at the sender's location A by observer Alice, while the other mode is open to local operations and
measurements in the receiver's location B, by observer Bob. Alice and Bob can communicate via a classical communication
channel. Thus Alice and Bob each have access to one of the two entangled sub systems described by
\begin{equation}
|{\cal E}\rangle_{AB}=\sqrt{(1-\lambda^2)}\sum_{n=0}^\infty \lambda^n|n\rangle_A\otimes|n\rangle_B
\label{squeezed}
\end{equation}
This state is generated from the vacuum state by the Unitary transformation
\begin{equation}
U(r)=e^{-r(a^\dagger b^\dagger - ab)}
\label{transform}
\end{equation}
where $\lambda=\tanh r$ and where $a,b$ refer to the mode accessible to Alice and the mode accessible to Bob respectively. 

  The entanglement of this state can be viewed in two ways. Firstly as an entanglement between quadrature phases in the two
modes (EPR entanglement) and secondly as an entanglement between number and phase in the two modes. We
can easily show that this state approximates the entanglement of an EPR state in the limit $\lambda\rightarrow 1$ or
$r\rightarrow\infty$. The quadrature phase entanglement is easily
seen by calculating the effect of the squeezing transformation Eq(\ref{transform}) in the Heisenberg picture. We first define
the quadrature phase operators for the two modes
\begin{eqnarray}
\hat{X}_A & = & a+a^\dagger\\
\hat{Y}_A & = & -i(a-a^\dagger)\\
\hat{X}_B & = & b+b^\dagger\\
\hat{Y}_B & = & -i(b-b^\dagger)
\label{qp}
\end{eqnarray}
Then
\begin{eqnarray}
Var(\hat{X}_A+\hat{X}_B) & = & 2e^{-2r}\\
Var(\hat{Y}_A-\hat{Y}_B) & = & 2e^{-2r}
\end{eqnarray}
where $Var(A)=\langle A^2\rangle-\langle A\rangle^2$ is the variance.
Thus in the limit of $r\rightarrow\infty$ the state $|{\cal E}\rangle$ approaches a simultaneous eigenstate of
$\hat{X}_A+\hat{X}_B$ and $\hat{Y}_A-\hat{Y}_B$. This is the analogue of the EPR state with position replaced by the real
quadratures $\hat{X}$ and the momentum replaced by the imaginary quadratures, $\hat{Y}$.  

This state is also entangled with respect to the correlation specified by the statement: {\em an equal number of photons in
each mode}. However it is not a perfectly entangled state, which would require the (unphysical) case of a uniform distribution
over correlated states. It can approach a perfectly entangled state of photon number asymptotically in the limit
$\lambda\rightarrow 1$. The reduced density operator of each mode is a thermal-like state with mean photon number
\begin{equation}
\bar{n}=\frac{\lambda^2}{1-\lambda^2}
\end{equation}
and thus the limit of a perfectly entangled state can only occur as the mean photon number goes to infinity, which is not
physical. For finite excitation, the distribution of correlated states is very close to uniform for values $n\ <\ e^{2r}$. This
suggests that in  practice this state can be used as a perfectly entangled state of photon number provided all other states
available have significant support on the photon number basis up to a maximum value of $n<<e^{2r}$. We now show that this is
indeed true if this state is used as a teleportation resource.

In the case of number and phase, it is obvious that the squeezed vacuum state is the zero eigenstate of the number difference
operator
\begin{equation}
\hat{J}_z=\frac{1}{2}(a^\dagger a-b^\dagger b)
\end{equation}
Not so obvious is the fact that as $\lambda\rightarrow 1$, the two modes become anti correlated in
phase. To see this we compute the canonical joint phase distribution for the two modes using the projection operator valued
measure
\begin{equation}
|\phi_A,\phi_B\rangle=\sum_{n,m=0}^\infty e^{in\phi_A+im\phi_B}|n\rangle_A\otimes|m\rangle_B
\end{equation}
normalised on $[-\pi,\pi]$ with respect to the measure $\frac{d\phi_Ad\phi_B}{4\pi^2}$. The joint distribution is
\begin{equation}
P(\phi_A,\phi_B)=\frac{1-\lambda^2}{|1-\lambda e^{i(\phi_A+\phi_B)}|^2}
\end{equation}
As $\lambda\rightarrow 1$ this distribution becomes very sharply peaked at $\phi_A=-\phi_B$. Thus the photon number in each
mode are perfectly correlated while the phase in each mode is highly anti correlated.

\section{Teleportation}
\subsection{Teleportation using a quadrature EPR state}
We first show how teleportation of continuous variables is possible using a perfect quadrature phase QND (quantum
nondemolition) measurement between two optical modes, A and B,  to create the entanglement resource.  The state that is
produced is an optical analogue of the EPR state discussed by Vaidman\cite{vaidman94}. Our presentation is completely
equivalent to that given by Vaidman, however we will use more conventional quantum optics notation. 

Consider the following entangled state of two modes A and B,
\begin{equation}
|X_1,P_1\rangle_{AB}=e^{-i\hat{Y}_A\hat{X}_B}|X_1\rangle_A\otimes|Y_1\rangle_B
\label{EPR}
\end{equation}
where the quadrature phase operators, $\hat{Y}_A\ ,\ \ \hat{X}_B$ are defined in Eq(\ref{qp}) and the states appearing in this
equation are the quadrature phase eigenstates,
\begin{eqnarray*}
\hat{X}_A|X_1\rangle_A & = & X_1|X_1\rangle_A\\
\hat{Y}_B|Y_1\rangle_B & = & Y_1|Y_1\rangle_B
\end{eqnarray*}
One then easily verifies that the state defined in Eq(\ref{EPR}) is a simultaneous eigenstate of $\hat{X}_A-\hat{X}_B$ and 
$\hat{Y}_A+\hat{Y}_B$ with respective eigenvlaues, $X_1, Y_1$.  The unitary transformation in Eq(\ref{EPR}) is generated by the
perfect QND Hamiltonian $H=\hat{Y}_A\hat{X}_B$, which realises a QND coupling between modes A and B. It is also the prototype
measurement coupling Hamiltonian first defined by von Neumann. It is important to realise that all perfect QND couplings are
a source of entanglement and a potential resource for teleportation. Needless to say this state is not a physical state, not
because the QND interaction cannot be achieved, but because the quadrature phase eigenstates appearing in Eq(\ref{EPR}) are
infinite energy states. However we can use arbitrary close approximations to these states given a sufficient energy resource,
as in the case of a squeezed vacuum state discussed below.  

In the protocol for teleportation based on this state, we now consider another mode, the target mode T, in an unknown state
$|\psi\rangle_T$. Joint quadrature phase measurements of $\hat{X}_T-\hat{X}_A$ and $\hat{Y}_T+\hat{Y}_A$  are made on modes T
and A, yielding two real numbers, $X_2,Y_2$ respectively. The total input state for the teleportation protocol is
\begin{equation}
|\Psi_{in}\rangle=|\psi\rangle_T\otimes|X_1,Y_1\rangle_{AB}
\end{equation}
The (unnormalised) conditional state of the total system after the measurement on A and T is given by the projection
\begin{equation}
|\tilde{\Psi}_{out}^{(X_2,P_2)}\rangle=\mbox{}_{AT}\langle X_2,Y_2|\psi\rangle_T|X_1,Y_1\rangle_{AB}\otimes|X_2,Y_2\rangle_{AT}
\end{equation}
Using the Eq(\ref{EPR}) we may write then write the conditional state of mode B as 
\begin{equation}
|\phi^{(X_2,P_2)}\rangle_B=\left [P(X_2,Y_2)\right ]^{-1/2}\hat{\Phi}(X_2,Y_2)|Y_1\rangle_B
\end{equation}
where 
\begin{equation}
P(X_2,Y_2)=\mbox{}_B\langle Y_1|\hat{\Phi}^\dagger\hat{\Phi}|Y_1\rangle_B
\end{equation}
is the  probability for the results $(X_2,Y_2)$. The state $|Y_1\rangle_B$ is an eigenstate of $\hat{Y}_B$ with eigenvalue
$Y_1$, which is determined by the initial choice of entangled state for A and B.  
The operator $\hat{\Phi}$, acts only on mode B
and  is defined by
\begin{equation}
\hat{\Phi}(X_2,Y_2)=\mbox{}_{AT}\langle X_2,Y_2|e^{-i\hat{Y}_A\hat{X}_B}|\psi\rangle_T\otimes|X_1\rangle_A
\end{equation}
Using the definition of the state $|X_2,Y_2\rangle_{AT}$,
\begin{equation}
|X_2,Y_2\rangle_{AT}=e^{i\hat{X}_A\hat{Y}_T}|X_2\rangle_T\otimes|Y_2\rangle_A
\end{equation}
where $\hat{X}_T|X_2\rangle_T=X_2|X_2\rangle_T$ and $\hat{Y}_A|Y_2\rangle_A=Y_2|Y_2\rangle_A$, it is possible to show that
\begin{equation}
|\psi^{(X_2,P_2)}\rangle_B=e^{iX_2Y_2}e^{iX_2\hat{Y}_B}e^{-iP_2\hat{X}_B}|\psi\rangle_B
\end{equation}
Thus up to a phase factor and two simple unitary transformations, the conditional state of B is the same as the initial unknown
state of the target T. If A now sends the results of the measurements $(X_2,Y_2)$ to the receiver, B, the phase factor and two
unitary transformations can be removed by local operations that correspond to a displacement in phase space by $X_2$ in the
real quadrature direction and $Y_2$ in the imaginary quadrature direction. The initial state of T has then been 'teleported'
to mode B at a distant location.

\subsection{Squeezed vacuum state teleportation using quadrature measurements}
In the introduction we noted that the squeezed vacuum state
\begin{equation}
|{\cal E}\rangle=e^{-r(a^\dagger b^\dagger - ab)}|0\rangle_{AB}
\end{equation}
is an approximation to the quadrature EPR state discussed in the previous section. In the limit of infinite squeezing, this
state becomes equivalent to the EPR state. We now show that the  two-mode squeezed vacuum state can be used for teleportation
with fidelity that approaches unity as the squeezing increases to infinity. 

We again assume perfect projective measurements
of the joint quadrature phase quantities, $\hat{X}_T-\hat{X}_A$ and $\hat{Y}_A+\hat{Y}_B$ on the target sate and the senders
part of the entangled mode, A, with the results $X,Y$ respectively. The (unnormalised) conditional state of total system after
the measurement is then seen to be given by
\begin{equation}
|\tilde{\Psi}^{(X,Y)}=\mbox{}_{T}\langle X|\otimes\mbox{}_A\langle Y|e^{i\hat{Y}_T\hat{X}_A}|\psi\rangle_T|{\cal
E}\rangle_{AB}
\end{equation}It is then easy to show that the state of mode B at the receiver is the pure state $|\phi_{XY}(r)\rangle_B$
with the wave function (in the $\hat{X}_B$ representation),
\begin{equation}
\phi_{XY}(x_1,x_2;r)=\int_{-\infty}^\infty dx_1dx_2e^{ix_1Y}{\cal G}(x_1,x_2;r)\psi(X-x_1)
\end{equation}
where $\psi(x)=\mbox{}_T\langle x|\psi\rangle_T$ is the wavefunction for the target state we seek to teleport. The kernel is
given by 
\begin{equation}
{\cal G}(x_1,x_2;r)=\frac{1}{\sqrt{2\pi}}\exp\left [-\frac{1}{4}(x_1+x_2)^2e^{2r}-\frac{1}{4}(x_1-x_2)^2e^{-2r}\right ]
\end{equation}

This state is clearly not the same as the state we sought to teleport. However in the limit of infinite squeezing,
$r\rightarrow\infty$, we find that ${\cal G}(x_1,x_2;r)\rightarrow \delta(x_1+x_2)$ and 
the state of mode B approaches
\begin{equation}
|\phi_{XY}(r)\rangle_B\rightarrow e^{iXY}e^{-iY\hat{X}_B}e^{iX\hat{Y}_B}|\psi\rangle_B
\end{equation}
which, up to the expected unitary translations in phase-space, is the required teleported state.

\subsection{Squeezed vacuum state teleportation using number and phase measurements}

In this section we explore to what extent teleportation is possible using the number phase entanglement implicit in the squeezed
vacuum state. In this case we expand the target state in the photon number basis as 
\begin{equation}
|\psi\rangle_T=\sum_{m=0}^\infty c_m |m\rangle_T
\end{equation}
Thus the input state to the receiver and sender is 
\begin{equation}
|\Psi_{in}\rangle=(1-\lambda^2)^{1/2}\sum_{n,m=0}^\infty \lambda^n c_m|m\rangle_T\otimes|n\rangle_A\otimes|n\rangle_B
\end{equation}
To facilitate the description of the joint measurements that need to be made on T and A modes at the receiver, we define the
eigenstates of the operator
\begin{equation}
\hat{J}_z=\frac{1}{2}(\hat{N}_T-\hat{N}_A)
\end{equation}
where $\hat{N}_T,\hat{N}_A$ are the number operators for modes T and A respectively. These eigenstates can be written as
pseudo angular momentum states as
\begin{equation}
\hat{J}_z|j,k\rangle_{AT}=k|j,k\rangle_{AT}
\label{angmom}
\end{equation}
where the eigenvlaue $j$ of $\hat{J}^2$ is determined by the result $\hat{J}^2=\frac{\hat{N}}{2}(\frac{\hat{N}}{2}+1)$ , where
$\hat{N}=\hat{N}_A+\hat{N}_T$ is the total photon number operator for modes $T$ and $A$ with eigenvlaue $N=0,1,2,\dots$. In
that case $j=\frac{N}{2}$. The relationship between the state Eq.(\ref{angmom}) and the original product number basis is
\begin{equation}
|j,k\rangle_{AT}=|j+k\rangle_T\otimes|j-k\rangle_A
\end{equation}
The combined state of the entire system may now be written
\begin{equation}
|\Psi_{in}\rangle=(1-\lambda^2)^{1/2}\sum_{j=0}^\infty\sum_{k=-j}^j \lambda^{j-k}c_{j+k}|j,k\rangle_{AT}\otimes|j-k\rangle_B
\end{equation}
Note that in this equation the sum over $j,k$ is over half integers as well as integers.

The teleportation protocol for number and phase requires that Alice make two measurements of a joint quantity on A and T. In
this case the first measurement will seek to determine one half the photon number difference as represented by $\hat{J}_z$,
while the second measurement will seek to determine the phase sum of the two modes. For the first measurements the possible
results are $k=\{0,\pm\frac{1}{2},\pm 1,\pm\frac{3}{2},\ldots\}$. Consider first the case of $k\ >\ 0$.  The conditional
(unnormalised) state of the entire system is 
\begin{equation}
|\Psi^{(k)}\rangle=(1-\lambda^2)^{1/2}\sum_{n=0}^\infty \lambda^nc_{n+2k}|n+2k\rangle_T\otimes|n\rangle_A\otimes|n\rangle_B
\end{equation}
where we have returned to the product number basis in preparation for the next measurement of phase sum. If the measurement
result was negative $k\ <\ 0$ , the conditional unnormalised state is
\begin{equation}
|\Psi^{(-k)}\rangle=(1-\lambda^2)^{1/2}\sum_{n=0}^\infty\lambda^{n+2k}c_n|n\rangle_T\otimes|n+2k\rangle_A\otimes|n+2k\rangle_B
\end{equation}

The {\em second} measurement is a measurement of the joint total phase operator for modes $T$ and $A$, defined by the
projection operator valued measure 
\begin{equation}
|\phi_+\rangle\langle \phi_+| =\sum_{n,m=0}^\infty \sum_{k=-min(n,m)}^{k=min(n,m)} |n,k\rangle_{AT}\langle k,m|
\end{equation}
Now, it must be said at once that such measurements are unphysical, however they do represent the limit of perfectly valid
(though rather impractical) discrete phase measurements\cite{MilWalls}.  As a result of
this measurement, Alice has a value $\phi_+$ for the phase. The corresponding conditional state of mode B, given a positive
number difference measurement is  
\begin{equation}
|\psi^{(k,\phi_+)}\rangle_B= \frac{(1-\lambda^2)^{1/2}}{\sqrt{P_+(k)}}\sum_{n=0}^\infty \lambda^n
c_{n+2k}e^{-i\phi_+(n+k)}|n\rangle_B
\end{equation}
while if a negative number difference result were obtained the state of mode B is
\begin{equation}
|\psi^{(-k,\phi_+)}\rangle_B=
\frac{(1-\lambda^2)^{1/2}}{\sqrt{P_-(-k)}}\sum_{n=0}^\infty\lambda^{n+2k}c_ne^{-i(n-k)\phi_+}|n+2k\rangle_B
\end{equation}
where $P(k)$ is in fact the probability for Alice to obtain the result $k$. This is given by
\begin{eqnarray}
P_+(k) & = & (1-\lambda^2)\sum_{n=0}^\infty\lambda^{2n}|c_{n+2k}|^2\\
P_-(-k) & = & (1-\lambda^2)\sum_{n=2k}^\infty\lambda^{2n} |c_{n-2k}|^2
\end{eqnarray}
with $k$ taken as positive in both equations.

Now it only remains for
Alice to communicate to Bob what value she got for the two measurements, that is the values
$k$ and $\phi_+$, and for Bob to find the appropriate conditional unitary transformations to reconstruct the state.  The phase
displacement part is quite straightforward. The receiver, B,  applies the local unitary transformation 
\begin{equation}
U(\pm k,\phi)=e^{i\phi(\hat{N_B}\pm k)}
\end{equation}
where $N_B$ is the number operator for the mode B. After this transformation the states become,
\begin{eqnarray}
|\psi^{(k)}\rangle & = & \frac{(1-\lambda^2)^{1/2}}{\sqrt{P_+(k)}}\sum_{n=2k}^\infty \lambda^{n-2k}c_n|n-2k\rangle_B\\
|\psi^{(-k)}\rangle & = & \frac{(1-\lambda^2)^{1/2}}{\sqrt{P_-(-k)}}\sum_{n=0}^\infty \lambda^{n+2k}c_n|n+2k\rangle_B
\end{eqnarray}
with $k>0$ in both cases. Naively one might think that  we can now apply a number
displacement operator, either up or down by
$2k$ , to reconstruct the state in an analogous fashion to the case of quadrature teleportation. While formally we can
construct such an operator (see below), there is going to be a problem with the case $k\ >\ 0$, as all the coefficients for
photon number less than $2k$ will be missing ! This result is directly attributable to the fact that the spectrum of the
number operator is bounded below by zero. We must accept this as a limit to teleportation when number phase measurements are
used and keep this in mind when trying to find more general teleportation schemes in the future.

What is the number displacement operator ? The generator of displacements for
number must be the canonical phase. Formally this is defined by 
\begin{equation}
{\cal D}(k)=\int_{-\pi}^\pi d\phi e^{ik\phi}|\phi\rangle\langle\phi|
\end{equation}
where
\begin{equation}
|\phi\rangle=\sum_{n=0}^\infty e^{in\phi}|n\rangle
\end{equation}
The fact that these basis states are not normalisable indicates that it is impossible in practice to realise a true number
displacement operator. However there are schemes that can reproduce arbitrarily well a number
displacement\cite{Dariano,Vitali}.

We first consider the example of the target state prepared in the number state $|N\rangle$. In this case the probability to
obtain a result $m$ for the measurement of the photon number difference operator $2\hat{J}_z$ on A and T is 
\begin{equation}
P(m) = \left\{\begin{array}{ll}
							        (1-\lambda^2)\lambda^{2(N-m)} & m\leq N\\
															0 & m\ >\ N
			             \end{array}\right .
\end{equation}
where $m=0,1,2,\ldots$.  The  most probable result is $m=N$, in which case the teleported state is the vacuum state
$|0\rangle_B$ which, given the data $m=N$ may be displaced back to $|N\rangle_B$, {\em independent of the value of} $\lambda$.
Indeed it is easy to see that we can teleport a number state perfectly regardless of the value of $\lambda$ provided that we
can make number displacements. This is in contrast to the quadrature case where fidelity does depend on $\lambda$. This is a
consequence of the perfect correlation between photon number for each mode in the squeezed vacuum state. However the
probabilities for different values of the photon number difference in A and T do depend on the value of $\lambda$. 

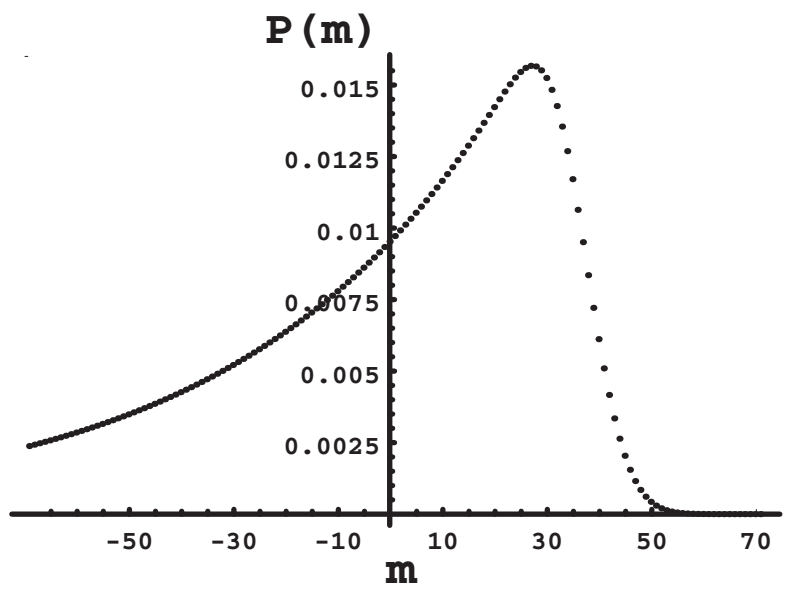
\begin{figure}
\caption{Figure 1: The probability distribution for obtaining a result $m$ for the number difference operator, $N_T-N_A$ for a
coherent state in the target with $\alpha=6.0,\lambda=0.99$}
\label{figone}
\end{figure}

Next consider the case of a coherent state $|\alpha\rangle$. This state has a Poisson photon number distribution with a mean of
$\bar{n}=|\alpha|^2$. The probability to observe a photon number difference $m$ between the target and the sender mode, A, is
\begin{equation}
P(m) = \left \{ \begin{array}{ll}\lambda^{-2|m|}(1-\lambda^2)e^{-|\alpha|^2(1-\lambda^2)}& m<0\\
																									(1-\lambda^2)e^{-|\alpha|^2}\sum_{n=0}^\infty \lambda^{2n}\frac{|\alpha|^{2(n+m)}}{(n+m)!}&m\geq 0
																														\end{array}\right .
\end{equation}
where $m=2k$ is an integer. 
This distribution is shown in figure \ref{figone}, with $\alpha=6, \lambda=0.99$. Note that the distribution is relatively flat
around $m=0$, that is around equal photon numbers in both A and T. It is easy to see that when $\lambda\rightarrow 1$, the
rapid fall off occurs for values $m>\bar{n}$. This is not too surprising as the most likely photon number in mode T is just
$\bar{n}$ and thus this is the largest possible value for the photon number difference between modes A and T. However the
minimum value for $m$ (which is negative) as determined by the largest photon number in mode A, which as $\lambda\rightarrow$
can be a large negative number. For this reason the distribution is highly asymmetric and falls off quite slowly for $m<0$.

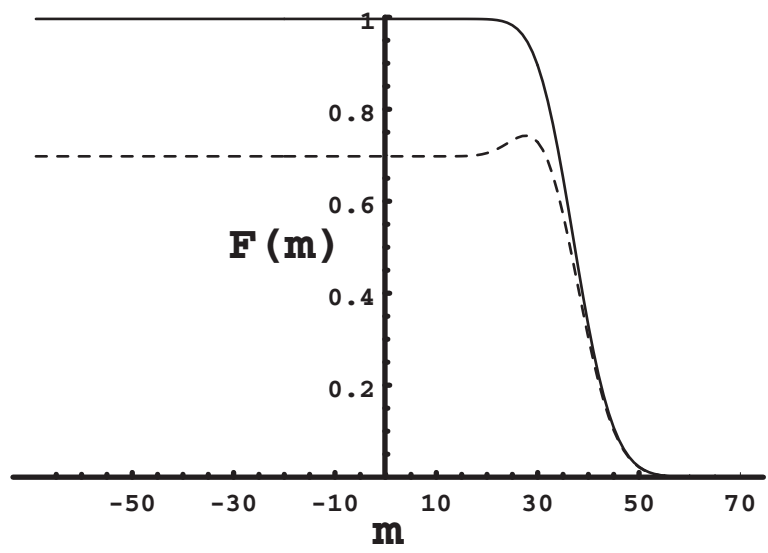
\begin{figure}
\caption{Figure 2: The fidelity versus the result $m$ for the number difference operator, $N_T-N_A$ for a
coherent state in the target with $\alpha=6,\lambda=0.9$ (dashed) and $\alpha=6.0,\lambda=0.99$ (solid).}
\label{figtwo}
\end{figure}

One performance measure for teleportation is the fidelity between the target state for mode B and the actual state teleported.
We will calculate the fidelity for the transported state after the appropriate number displacement operator has acted.
This is defined by 
\begin{equation}
F(m)=|\mbox{}_B\langle\psi|\tilde{\psi}^{(m)}\rangle_B|^2
\end{equation}
where $|\tilde{\psi}^{(m)}\rangle_B$ is the teleported  and displaced state at the receiver, B, given a photon number
difference measurement result, $m$, at the sender, A and T. The fidelity is given by
\begin{equation}
F(m)=\left \{\begin{array}{ll}
							\frac{(1-\lambda^2)}{P_+(m)}e^{-2|\alpha|^2}\left |\sum_{n=0}^\infty\lambda^n\frac{|\alpha|^{2(n+m)}}{(n+m)!}\right |^2
&m\geq 0\\
\exp\left[-|\alpha|^2(1-\lambda)^2\right] & m<0
\end{array}\right .
\end{equation}
The fidelity is plotted in figure \ref{figtwo} for $\alpha=6$ and two values of $\lambda$. We see that for $\lambda\rightarrow
1$ the fidelity is very close to unity until there is a chance of obtaining a positive photon number difference which exceeds
the average photon number in the target state we wish to teleport. However we see from figure \ref{figone} that this is likely
to happen with rapidly decreasing probability.

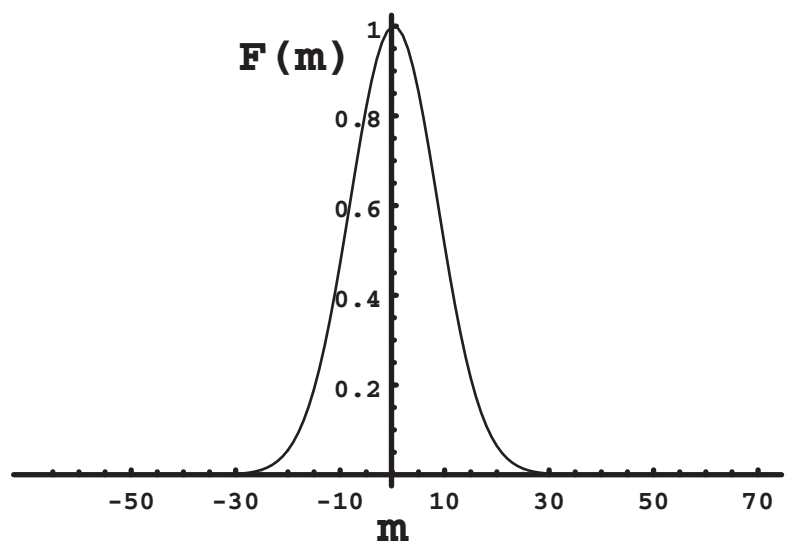
\begin{figure}
\caption{Figure 3: The fidelity versus the result $m$ for the number difference operator, $N_T-N_A$ when no attempt is made to
dispalce the teleported state conditioend in this result, for a coherent state in the target with $\alpha=6,\lambda=0.9$
(dashed)}
\label{figthree}
\end{figure}

Given the current difficulty of realising a photon number displacement operator it is of interest to determine the fidelity
when no attempt is made to displace the final state. If we assume that the target state is a coherent state with amplitude
$\alpha$, the fidelity when the results of the photon number difference measurement is zero, $m=0$, is
\begin{equation}
F(0)=e^{-|\alpha|^2(1-\lambda)^2}
\end{equation}
If we note that the mean photon number in the entanglement resource shared between A and B is just that for a squeezed vacuum
state , $\bar{n}_{SV}=\lambda^2/(1-\lambda^2)$, we may write the fidelity as 
\begin{equation}
F(0)=\exp\{-\frac{\bar{n}}{\bar{n}_{SV}}\lambda^2\}
\end{equation}
This indicates that when the mean photon number in the entanglement resource is significantly greater than that in the target
state, the teleportation is high fidelity. Indeed in the limit that $\lambda\rightarrow 1$, the teleportation for a result
$m=0$ is perfect.   Of course the fidelity falls off if $m\neq 0$ unless we act with the number displacement operator to shift
 the received state.  If we do not (or possibly cannot) do that the fidelity falls of in a Gaussian like fashion, which for
$|\alpha|>>1$ has a width that scales like half the mean photon number in the target state, $\bar{n}/2$. This is shown in
figure \ref{figthree}. 
	
\section{Conclusion}
We have shown how the imperfect  entanglement of a two-mode  squeezed vacuum state can be used for teleportation of an unknown
quantum state for two different measurement protocols at the sender. One protocol is based on quadrature phase measurements
and is suggested by the fact that a squeezed vacuum state is an approximation to an EPR correlated state for quadrature phase
amplitude variables. However a squeezed vacuum state is also entangled with respect to photon number difference and phase sum
in the two modes. This suggests a protocol based on number and phase measurements at the sender. While such measurements are
just beyond the reach of current experiments in quantum options, our examples suggest that a given entanglement resource admits
more than one teleportation protocol. In the case of a squeezed vacuum state the quadrature phase protocol is simpler based on
current technology. However this may not be true for other entanglement resources, or other realisations of the entanglement.
In fact any perfect QND interaction between two systems is a potential entanglement resource and determining the best
teleportation protocol may be a non trivial exercise.

\begin{acknowledgments}
We would like to thank the Benasque Centre for Physics for support during the visit at which this paper was written,
and also W. Munro for useful discussions. 
\end{acknowledgments}

\end{document}